\documentclass[onecolumn]{IEEEtran}
\usepackage{cite}
\usepackage{amsmath,amssymb,amsfonts}
\usepackage{algorithmic}
\usepackage{graphicx}
\usepackage{textcomp}
\usepackage{xcolor}
\usepackage{enumerate}
\usepackage{multirow}
\usepackage{caption}
\usepackage{placeins}
\newcommand{\RomanNumeralCaps}[1]
    {\MakeUppercase{\romannumeral #1}}
\def\BibTeX{{\rm B\kern-.05em{\sc i\kern-.025em b}\kern-.08em
    T\kern-.1667em\lower.7ex\hbox{E}\kern-.125emX}}
    \usepackage{balance}
    \usepackage{flushend}
    \usepackage{epstopdf}
\usepackage{amsmath}
\DeclareMathOperator{\diag}{diag}
\usepackage{array}
\usepackage{nicefrac}
\usepackage[font=footnotesize]{caption}
\makeatletter
\renewenvironment{abstract}{
    \centerline{\bfseries Abstract}
    \vspace{0.5em}
}{}
\makeatother

\begin{document}
\title{Design Considerations Based on Stability for a Class of TCP Algorithms
\\}
\author{\IEEEauthorblockN{Sreekanth Prabhakar and Gaurav Raina}\\
\IEEEauthorblockA{Department of Electrical Engineering, Indian Institute of Technology Madras, Chennai 600036, India}\\
ee20d005@smail.iitm.ac.in, gaurav@ee.iitm.ac.in}
\maketitle
\thispagestyle{plain}
\pagestyle{plain}
\begin{abstract}
Transmission Control Protocol (TCP) continues to be the dominant transport protocol on the Internet. The stability of fluid models has been a key consideration in the design of TCP and the performance evaluation of TCP algorithms. Based on local stability analysis, we formulate
some design considerations for a class of TCP algorithms. We begin with deriving sufficient conditions for the local stability of a generalized TCP algorithm in the presence of heterogeneous round-trip delays. Within this generalized model, we consider three specific variants of TCP: TCP Reno, Compound TCP, and Scalable TCP. The sufficient conditions we derive are scalable across network topologies with one, two, and many bottleneck links. We are interested in networks with intermediate and small drop-tail buffers as they offer smaller queuing delays. The small buffer regime is more attractive as the conditions for stability are decentralized.  TCP algorithms that follow our design considerations can provide stable operation on any network topology, irrespective of the number of bottleneck links or delays in the network. 
\end{abstract}
\section{Introduction}
Transmission Control Protocol (TCP) is responsible for end-to-end congestion control on the Internet. Over the past decades, numerous TCP algorithms with different design principles catering to diverse applications have been proposed (see \cite{TCP_survey,kelly_scalable,ren_mmwave,tcp_4g5g,tcp_ml}). Among these algorithms, those represented by an underlying mathematical model such as the fluid model \cite{Misra} offer additional feasibility to control-theoretic analyses (see \cite{control_red,Improved_control,Paganini,Glenn_1,Glenn_2}). TCP algorithms often compromise their stability, as they resort to aggressive window growth functions to meet the rising demands of the users and the network. Studies have shown that instabilities arising from protocol design can significantly impact network performance\cite{large_mux}. Hence it is important to consider the stability aspects when formulating design considerations for TCP algorithms.\par
 Stability is an essential characteristic of any control algorithm. In \cite{Kelly}, the authors presented two classes of congestion control algorithms, namely primal and dual algorithms, which were shown to be asymptotically stable. Sufficient conditions for the local stability of primal algorithms under the assumption of homogeneous round-trip delays were derived in \cite{Tan_Johari}. Based on the simulation results, the authors also presented a conjecture that these sufficient conditions may also hold in the presence of heterogeneous delays. This conjecture was verified to be true in \cite{Glenn_1} and similar conditions for a TCP-like algorithm were derived in \cite{Glenn_2}. The stability and performance of TCP algorithms are affected by the buffer sizes equipped at the network routers. The research community has pointed out that the rule of thumb for buffer sizing, which states that the buffer size should be equal to the bandwidth-delay product (BDP), is outdated and unsuitable for core routers (see \cite{Appenzeller,part_1,part_2,part_3,buffersize_update}). In \cite{large_mux}, the authors considered large, intermediate, and small buffer regimes. They showed that different buffer sizes gave rise to different queueing models and small buffers aid stability. Studies in \cite{part_2} also found that large buffers lead to instability. Although an Active Queue Management (AQM) \cite{aqm_survey, rate_vs_queue} may be deployed to improve stability, it will increase computational costs at the routers and may involve complex parameter tuning \cite{aweya_controltheory}. An alternative approach is to work with intermediate and small buffer regimes which offer smaller queueing delays. Given the increasing demand for latency-sensitive applications, we find these buffer regimes attractive for future networks.\par
 TCP/AQM systems studies mostly focus on topologies with a single bottleneck link (see \cite{lin_tcp_eval,elasticTCP}). Most works that include multiple bottleneck links, consider topologies with only two bottleneck links \cite{SRQPI_bisoy} or a set of bottleneck links in tandem \cite{Ma_backstepping}. Analytical studies that deal with multiple bottleneck links often allow for only a specific variant of TCP \cite{comb_bufferbloat}. On a large network such as the Internet, users may use distinct variants of TCP, depending on their application or the host operating system. The models and scenarios chosen in the studies should accommodate the diverse traffic mix, delays, and network topologies.\par
We formulate some design considerations based on stability for a class of TCP algorithms. First, we derive sufficient conditions for the local stability of networks operating a class of TCP algorithms in the presence of heterogeneous round-trip delays. Our analysis considers a generalized TCP model. Within this generalized model, we consider three specific variants of TCP: TCP Reno, Compound TCP \cite{compound_tcp}, and Scalable TCP \cite{kelly_scalable}.  The sufficient conditions we derive are scalable across network topologies with one, two, and many bottleneck links. We are interested in networks with intermediate and small drop-tail buffers as they offer smaller queueing delays. The small buffer regime is more attractive as the conditions for stability are decentralized. As our network model accommodates heterogeneous round-trip delays, multiple bottlenecks, and multiple variants of TCP, the insights we gather are useful in designing new TCP algorithms for the real world. TCP algorithms that follow our design considerations can offer stable operation on any network topology, irrespective of the number of bottleneck links or round-trip delays in the network.\par
The rest of the paper is organized as follows. In section \RomanNumeralCaps{2}, we present the models we use for TCP sources and the packet-dropping probability at the routers. We derive the sufficient conditions for local stability of networks with a single bottleneck link in section \RomanNumeralCaps{3}. In section \RomanNumeralCaps{4}, we consider networks with two bottleneck links in tandem. A topology with two edge routers feeding a core router is considered in \RomanNumeralCaps{5}. The stability of a generalized network topology with an arbitrary number of bottleneck links is analyzed in section \RomanNumeralCaps{6}. In sections \RomanNumeralCaps{7} and \RomanNumeralCaps{8}, we present our observations on how the presence of multiple TCP variants and heterogeneous delays impact the stability of networks. A design aspect that can help TCP algorithms improve their performance without compromising their stability is presented in section \RomanNumeralCaps{9}. We conclude our paper in section \RomanNumeralCaps{10}.
\section{Models}
Let $S$ be the set of all sources (users) and $L$ be the set of all links in the network. Each source $j$ is connected to a distinct destination node through a unique route. Let $R_j$ be the set of bottleneck links along the path of user $j$, $R_j\subset L$. The sources employ a generalized class of TCP algorithms in which each source $j$ increases its congestion window $w_j(t)$ by $i_j(w_j(t))$ for every acknowledgment (ACK) and decreases $w_j(t)$ by $d_j(w_j(t))$ for every loss. The fluid model equation for a source using generalized TCP \cite{part_2} is given by
\begin{equation}
\begin{aligned}
    \frac{dw_j(t)}{dt}\! =\hspace{0.1cm}&\frac{w_j(t-T_j)}{\ T_j}\! \biggl(\! i_j\big(w_j(t)\big)\big( 1 - q_j(t) \big)- d_j\big(w_j(t)\big)  q_j\big(t\big) \!\biggr),\label{fluid_eqn}
\end{aligned}
\end{equation}
where $T_j$ denotes the average round-trip time (RTT) of source $j$ and $q_j(t)$ represents the aggregate loss probability inferred by source $j$. The average sending rates may be expressed as
\begin{equation}
    x_j(t)=\frac{w_j(t)}{T_j}.
\end{equation}
Writing \eqref{fluid_eqn} in terms of the average sending rate, we get
\begin{equation} \label{eqn1}
\begin{aligned}
   T_j \frac{dx_j(t)}{dt} =\hspace{0.1cm}& x_j(t-T_j)i_j(x_j(t))- x_j(t-T_j)q_j(t)(i_j(x_j(t)+d_j(x_j(t)). 
\end{aligned}
\end{equation}
 Let $p_l(t)$ denote the instantaneous packet drop probability at link $l$, then $q_j(t)$ is given by 
\begin{equation} \label{eqn2}
\begin{aligned}
    q_j(t)&=1-\prod_{l \in R_j}^{}\biggl(1-p_l(t-(T_j-T_{jl})) \biggr),
\end{aligned}
\end{equation}
where $T_{jl}$ is the forward delay from source $j$ to the link $l$. The total arrival rate  at link $l$ can be expressed as
\begin{equation}
    \begin{aligned}
        y_l(t)=\sum\limits_{k:l \in R_k}^{}x_k(t-T_{kl}).
    \end{aligned}
\end{equation}We assume that the network routers are equipped with small or intermediate drop-tail buffers. For small drop-tail buffers with smooth traffic, the packet drop probability may be approximated  as \cite{large_mux} 
\begin{equation}\label{p_d}
    p_l(t)=\left ( \frac{y_l(t)}{C_l} \right )^{B_l},
\end{equation}
where $B_l$ and $C_l$ are the buffer size and capacity associated with link $l$. For bursty traffic, the packet drop probability for small drop-tail buffers \cite{burstyTCP} is given by 
\begin{equation}
    p(t)=\Biggl(\frac{\sum\limits_{k \in S}^{}x_j(t-T_{j1})}{\frac{C}{M}}\Biggr)^{\scalebox{1}{$\frac{B}{M}$}},
\end{equation}
where $M$ is the average burst size (pkts). 
In the case of an intermediate buffer with drop-tail policy, the packet drop probability\cite{large_mux} is given by
\begin{equation}\label{p_d_i}
    p_l(t)=\left ( 1-\frac{C_l}{y_l(t)} \right )^{+},
\end{equation}
where $(f)^+$ is defined as $max(f,0).$
\section{Topology 1: A single Bottleneck Link}
\begin{figure}[h]
    \centering
    \includegraphics[width=8cm]{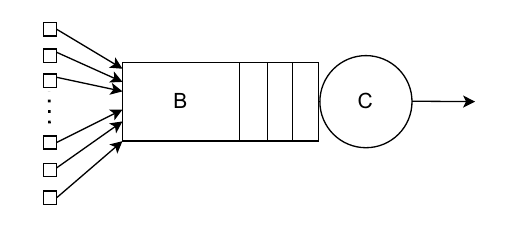}
    \caption{$n$ sets of flows sharing a single bottleneck link with capacity $C$ and buffer size $B$. Each set is characterized by its window adjustment functions ($i_j(w(t)), d_j(w(t))$), average sending rate ($x_j(t)$) and RTT ($T_j$). }
    \label{fig1}
\end{figure}

In this section, we derive sufficient conditions for the local stability of networks with a single bottleneck link. Consider $n$ sets of flows sharing a single bottleneck link with capacity $C$ and buffer size $B$ (see Fig.\ref{fig1}). Each set is characterized by its window adjustment functions ($i_j(w_j(t)), d_j(w_j(t))$), average sending rate $x_j(t)$ and average RTT of $T_j$. Let $p(t)$ be the packet-dropping probability at the bottleneck link at time $t$. 
\FloatBarrier
The fluid model equation for source $j$ is
\begin{equation}\label{1b_1d}
\begin{aligned}
   T_j \frac{dx_j(t)}{dt}= & x_j(t-T_j)i_j(x_j(t))- x_j(t-T_j)p(t-T_j+T_{j1})(i_j(x_j(t))+d_j(x_j(t))). 
\end{aligned}
\end{equation}
We linearize \eqref{1b_1d} about the equilibrium point given by
\begin{equation}
    \frac{i_j^*}{i_j^*+d_j^*}=p^*,
\end{equation}
where $f^*$ denotes the equilibrium value of the function $f(t)$. Let $u_j(t)$ be an arbitrary perturbation from the equilibrium rate i.e., $u_j(t)=x_j(t)-x_j^*$, where $x_j^*$ is the equilibrium sending rate. The linearized equations are
\begin{equation} \label{eqn4}
    \begin{aligned}
        T_j\frac{du_j(t)}{dt}=&-a_ju_j(t)-b_ju_j(t-T_j)-\sum\limits_{k\in S,k\neq j}^{}c_{jk}u_k(t-T_j-(T_{k1}-T_{j1})).
    \end{aligned}
\end{equation}
The coefficients are given by
\begin{equation}
    \begin{aligned}
        a_j&=-x_j^*i_j'+x_j^*p^*(i_j'+d_j')
    \end{aligned}
\end{equation} and 
\begin{equation}
    \begin{aligned}
    \hspace{-1cm}b_j&=-x_j^*\frac{p' i_j^*}{p^*}=c_{jk}.
    \end{aligned}
\end{equation}
Here $i_j'$ and $d_j'$ respectively, are equilibrium values of the derivatives of $i_j(x(t))$ and $d_j((x(t))$ with respect to $x_j(t)$. Also, $p'$ represents the equilibrium value of the derivative of the packet drop probability w.r.t. an associated source rate. Equation \eqref{eqn4} can be represented in the Laplace domain as
\begin{equation}\label{eqn14}
    \begin{aligned}
        u_j(s)=&\frac{e^{-sT_j}}{a_j+sT_j}\biggl(\frac{p'i_j^*x_j^*}{p^*}\Big(u_j(s)-\sum\limits_{k\in S,k\neq j}^{}e^{-s(T_{k1}-T_{j1})}u_k(s)\Big)\biggr).
    \end{aligned}
\end{equation}
\subsection{Intermediate drop-tail buffers}
In the case of intermediate drop-tail buffers, the packet drop probability is given by 
\begin{equation}
    p(t)=\Biggl(1-\frac{C}{\sum\limits_{k \in S}^{}x_j(t-T_{j1})}\Biggr)^+.
\end{equation}
Hence, equation \eqref{eqn14} can be written as 
\begin{equation}\label{eqn26}
   \begin{aligned}
          u_j(s)=&\frac{e^{-sT_j}}{a_j+sT_j}\biggl(\frac{(1-p^*)}{p^*} \frac{i_j^*x_j^*}{y^*}\Big(u_j(s)-\sum\limits_{k\in S,k\neq j}^{}e^{-s(T_{k1}-T_{j1})}u_k(s)\Big)\biggr).  
    \end{aligned}
\end{equation}
\par In \cite{Glenn_2}, the author derived the sufficient conditions for local stability for a similar set of equations, by bounding the eigenvalues of loop transfer functions. The work in \cite{Glenn_2} considered only a specific set of window incrementing and decrementing functions for TCP algorithms. Since we use generalized functions $i(x(t))$ and $d(x(t))$, our model may readily be applied to a broad class of TCP algorithms that can be represented using the fluid model \eqref{fluid_eqn}. Let the connectivity matrix $R(s)$ be defined as 
\begin{equation}
    R_{lj}(s)=\left\{\begin{matrix}
e^{-sT_{jl}} & \textnormal{if }l\in R_{j} \\ 
0 & \textnormal{otherwise}.
\end{matrix}\right.
\end{equation}
For the single bottleneck link, $R(s)$ is given by
\begin{equation}
    R(s)=[e^{-sT_{11}} \textnormal{       } e^{-sT_{21}} \textnormal{ ...       } e^{-sT_{N1}}].
\end{equation}
Let $\bar u=[u_1(s) \textnormal{  } u_2(s) \textnormal{ ...       }  u_N(s)]^T$. The loop transfer function from $\bar{u}$ to itself, known as the return ratio, is given by
\begin{equation}\label{L_s_i}
\begin{aligned}
    L(s)=&\diag\Biggl(\frac{e^{-sT_j}}{a_j+sT_j}\Biggr) \diag\biggl(\frac{(1-p^*)i_{j}^*}{p^*}\biggr) \diag \Biggl(\frac{x_j^*}{y^*}\Biggr) R^T(-s)R(s).
    \end{aligned}
\end{equation}
\par
\begin{figure}[h]
    \centering
    \includegraphics[width=0.65\linewidth]{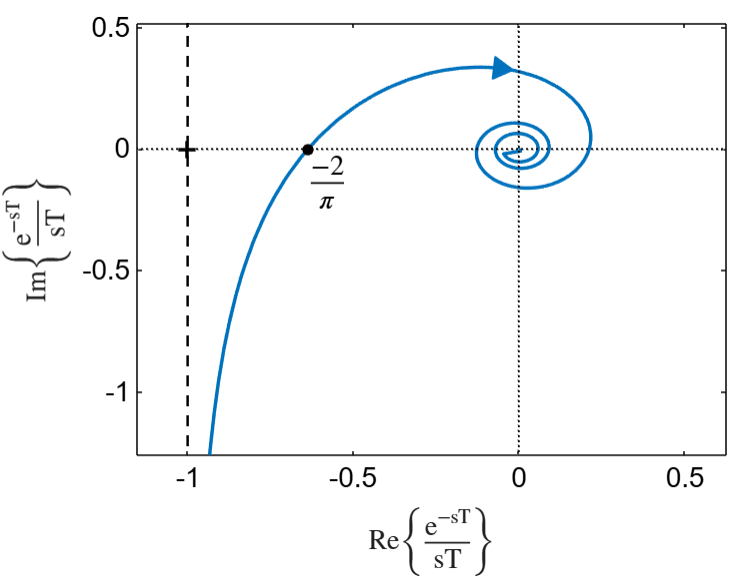}
    \caption{Nyquist plot of $\frac{e^{-sT}}{sT}$. The Nyquist plot cuts the real axis at $\frac{-2}{\pi}$. An additional gain of $\frac{\pi}{2}$ can be permitted for the system without compromising stability.}
    \label{nyquist_plot}
\end{figure}

 From control theory, we know that the Nyquist plot of stable transfer functions should not encircle (-1,0). The factor $\dfrac{e^{-sT_j}} {a_j+sT_j}$ \vspace{0.1cm}is stable for $a_j>0$ and the corresponding Nyquist plot cuts the real axis to the right of (-1,0) (see Fig.\ref{nyquist_plot}). It has been shown in \cite{Glenn_2} that to ensure the stability of the entire loop transfer function, we need to limit the magnitude of its eigenvalues. Let us use the maximum absolute row sum of the matrix as an upper bound for its eigenvalues. Let $\rho(A)$ denote the spectral radius of $A$. Then,
\FloatBarrier
\begin{equation}
    \begin{aligned}\rho\Biggl(\diag\Biggl(\frac{x_j^*}{y^*}\Biggr)R^T(-s)R(s)\Biggr)&=\rho\Biggl(R^T(-s)\hspace{0.1cm}R(s)\hspace{0.1cm}\diag\Biggl(\frac{x_j^*}{y^*} \Biggr)\Biggr )\\
    &=  \left\|R^T(-s)\right\|_\infty.\left\|R(s)\hspace{0.1cm}\diag\Biggl(\frac{x_j^*}{y^*}\Biggr) \right\|_\infty\\
    &=1,
    \end{aligned}
\end{equation}
since $y_l^*=\sum\limits_{j: l\in R_j} x_j^*$. The only remaining term in the return ratio \eqref{L_s_i} is $\diag\biggl(\dfrac{(1-p^*)i_{j}^*}{p^*}\biggr)$. Hence, the system will be stable if each TCP source or user satisfies the condition
\begin{equation}\label{sc_1_i}
\begin{aligned}
    \frac{i_j(w_j(t))(1-p^*)}{p^*}&<\frac{\pi}{2} \hskip 2em  \textnormal{for } j \in S.
\end{aligned} 
\end{equation}
The Nyquist plot of $\dfrac{e^{-sT_j}}{a_j+sT_j}$ cuts the real axis to the right of $\dfrac{2}{\pi}$ when $a_j > 0$. Thus a maximum gain of $\dfrac{\pi}{2}$ can be permitted without compromising stability. \par
Thus, \eqref{sc_1_i} gives sufficient conditions for the local stability of a network having a single bottleneck link with an intermediate drop-tail buffer in the presence of heterogeneous delays. We can see that for small values of $p^*$, this condition is too difficult to maintain. 
\subsection{Small drop-tail buffers }
\subsubsection{Smooth traffic}
For small drop-tail buffers with smooth traffic, the packet drop probability is given by 
\begin{equation}
    p(t)=\Biggl(\frac{\sum\limits_{k \in S}^{}x_j(t-T_{j1})}{C}\Biggr)^B.
\end{equation}
Thus \eqref{eqn14} may be written as 
\begin{equation}\label{eqn17}
    \begin{aligned}
          u_j(s)=&\frac{e^{-sT_j}}{a_j+sT_j}\biggl(\frac{Bi_j^*x_j^*}{y^*}\Big(u_j(s) -\sum\limits_{k\in S,k\neq j}^{}e^{-s(T_{k1}-T_{j1})}u_k(s)\Big)\biggr).  
    \end{aligned}
\end{equation}
The return ratio is given by
\begin{equation}\label{L_s}
\begin{aligned}
L(s)= &\diag\Biggl(\frac{e^{-sT_j}}{a_j+sT_j}\Biggr) \diag(Bi_{j}^*)\diag \Biggl(\frac{x_j^*}{y^*}\Biggr)R^T(-s)R(s).
\end{aligned}   
\end{equation}
Proceeding with the control-theoretic analysis, we get sufficient conditions for local stability in the case of small drop-tail buffers with smooth traffic as
\begin{equation}
 \hspace{-2cm}   Bi_j^*< \frac{\pi}{2}
\end{equation}
or
\begin{equation}\label{sc_1}
\begin{aligned}
    i_j(w_j(t))&<\frac{\pi}{2B} \hskip 2em  \textnormal{for } j \in S.
\end{aligned} 
\end{equation}
 For small values of packet-drop probability ($p^*<<1$), we see that as we move from intermediate buffers to small buffers, sufficient conditions for stability become more relaxed.\\
\subsubsection{Bursty traffic}
For small drop-tail buffers with bursty traffic, the packet drop probability \cite{burstyTCP} is given by 
\begin{equation}
    p(t)=\Biggl(\frac{\sum\limits_{k \in S}^{}x_k(t-T_{k1})}{\frac{C}{M}}\Biggr)^{\scalebox{1}{$\frac{B}{M}$}},
\end{equation}
where $M$ is the average burst size (pkts). Subsequent analysis yields sufficient conditions for local stability as
\begin{equation}
   \hspace{-2cm}\frac{B}{M}i_j^*< \frac{\pi}{2},
\end{equation}
or
\begin{equation}
\begin{aligned}
    i_j(w_j(t))&<\frac{\pi M}{2B} \hskip 2em  \textnormal{for } j \in S.
\end{aligned} 
\end{equation}
\par The generalized TCP model \eqref{fluid_eqn} caters to a wide class of TCP algorithms. We consider three variants of TCP in particular: TCP Reno, Compound TCP, and Scalable TCP. Sufficient conditions for the local stability of these algorithms with intermediate and small drop-tail buffers are given in Table \ref{conditions_topology_1}.
The results show that for small values of packet loss probability ($p^*<<1$), the sufficient conditions we obtain are more stringent for intermediate-sized buffers. 
\begin{table}[h]
    \small
    \captionsetup{justification=justified} 
    
    \caption{\small{Sufficient conditions for local stability of networks with single bottleneck link}}
    \centering
\begin{tabular}{|p{2.5cm}|>{\centering\arraybackslash}p{3.5cm}|>{\centering\arraybackslash}p{3.7cm}|}
    \hline
 \centering \vspace{0.3cm}Topology & \multicolumn{2}{c|}{\vspace{-0.6cm}}\\ 
     & \multicolumn{2}{c}{Sufficient conditions for local stability \vspace{0.05cm}}\\
    \cline{2-3}
   & \centering{\vspace{0.02cm}Small drop-tail buffer} &\vspace{0.02cm}Intermediate drop-tail buffer \\ 
   \hline
   \vspace{0.05cm} Generalized TCP & \vspace{0.05cm} $i_j(w_j(t))<\dfrac{\pi}{2B}$ & \vspace{0.05cm}$i_j(w(t))<\dfrac{\pi}{2}\dfrac{q_j^*}{(1-q_j^*)}$ \vspace{0.05cm}\\ 
   \vspace{0.05cm} TCP Reno &  \vspace{0.05cm} $\dfrac{1}{w_j}<\dfrac{\pi}{2B}$&\vspace{0.05cm}$ \dfrac{1}{w_j} < \dfrac{\pi}{2}\dfrac{q_j^*}{(1-q_j^*)}$ \vspace{0.05cm}\\ 
   
    \vspace{0.05cm}Compound TCP & \vspace{0.05cm}$\alpha w_j^{k-1}<\dfrac{\pi}{2B}$ &\vspace{0.06cm}$\alpha w_j^{k-1}<\dfrac{\pi}{2} \dfrac{q_j^*}{(1-q_j^*)}$ \vspace{0.05cm}\\

    \vspace{0.05cm}Scalable TCP &  \vspace{0.05cm}$a<\dfrac{\pi}{2B}$ &\vspace{0.05cm} $a<\dfrac{\pi}{2}\dfrac{q_j^*}{(1-q_j^*)}$ \\ \vspace{0.05cm} &\vspace{0.05cm}& \vspace{0.05cm}\\
    \hline
\end{tabular}
\label{conditions_topology_1}
\end{table}
\FloatBarrier
\section{Topology 2: Two Bottleneck Links in Tandem}
\begin{figure}[!h]
    \centering
    \includegraphics[width=12cm,height=3cm]{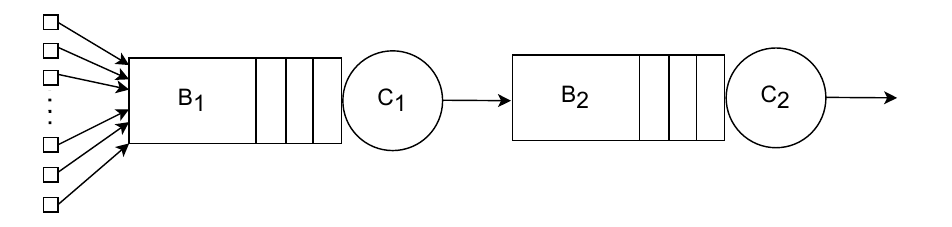}
    \caption{Two bottleneck links in tandem shared by $n$ flows. The links have capacities $C_1$, $C_2$ and associated buffer sizes $B_1$, $B_2$ respectively. Each set of flows is characterized by its average sending rate ($x_j(t)$), window adjustment functions ($i_j(x_j(t)), d_j(x_j(t))$), and RTT ($T_j$).}
    \label{fig2}
\end{figure}

Consider $n$ flows with heterogeneous round-trip delays passing through two bottleneck links $l_1$ and $l_2$ connected in series as shown in Fig. \ref{fig2}.
\FloatBarrier
The connectivity matrix for the topology is given by 
\begin{equation}
 R(s)=   \begin{bmatrix}
 e^{-sT_{11}}& ... &e^{-sT_{N1}}  \\
e^{-sT_{12}} & ... &e^{-sT_{N2}}  \\
\end{bmatrix}.
\end{equation}
Note that $T_{j2}=T_{j1}+\delta$, where $\delta$ is the time taken by a packet to traverse the first link. This makes the two rows of $R(s)$ proportional to each other, which implies that only one of them will act as a bottleneck in the equilibrium state \cite{Paganini}. Hence, the above scenario eventually reduces to the single bottleneck case.
\subsection{Intermediate drop-tail buffers }
Sufficient conditions for stability remain as 
\begin{equation}
\begin{aligned}
    \frac{i_j(w_j(t))(1-p^*)}{p^*}&<\frac{\pi}{2} \hskip 2em \textnormal{for } j \in S,
\end{aligned} 
\end{equation}
where $p^*$ refers to the equilibrium packet drop probability at the link with the smallest capacity.
\subsection{Small drop-tail buffers}
Sufficient conditions for stability are given by 
\begin{equation}
    i_j(w_j(t))<\frac{\pi}{2B} \hskip 2em  \textnormal{for } j \in S,
\end{equation}
where $B$ is the buffer size associated with the link having the least capacity.
\section{Topology 3: Two Edge routers feeding a core router}
\begin{figure}[h]
    \centering
    \includegraphics[width=12cm,height=6cm]{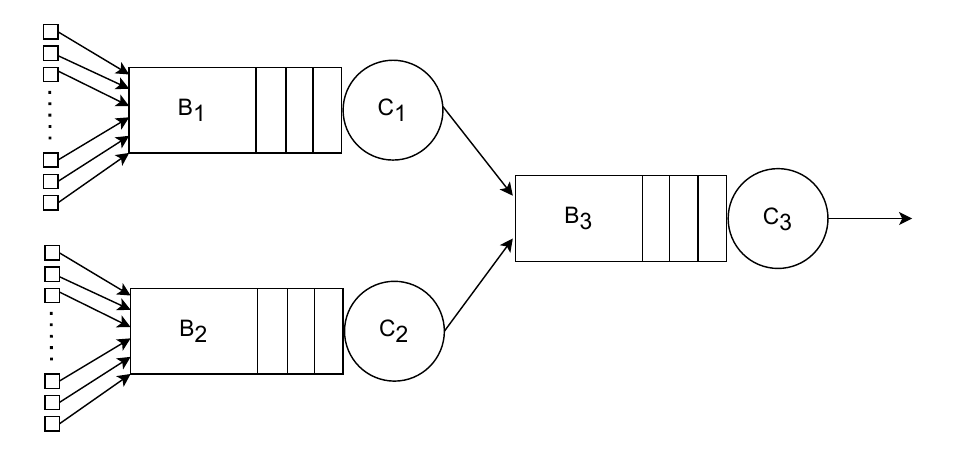}
    \caption{Two edge routers feeding into a core router: TCP flows with heterogeneous round-trip delays entering two edge routers with capacities $C_1$ and $C_2$, respectively. Each set is characterized by its window adjustment functions (i.e., $i_j(x_j(t))$ and $d_j(x_j(t))$), RTT ($T_j$) and average sending rate ($x_j(t)$). $B_1$ and $B_2$ are the buffer sizes associated with the edge routers. $C_3$ and  $B_3$ respectively denote the link capacity and buffer size associated with the core router.}
    \label{fig3}
\end{figure}

Consider a network where two edge routers feed a core router as shown in Fig. \ref{fig3}. 
\FloatBarrier
Each flow is characterized by its average sending rate and window adjustment functions. Let $S$ be the set of all sources and $S_m$ be the set of sources entering the $m^{th}$ edge router. The fluid model equations for the flows are given by \eqref{eqn1}. The aggregate loss probabilities for the flows may be expressed as
\begin{equation}
\begin{aligned}
        q_j(t)&=1- \big(1-p_m(t-T_j+T_{jm})\big)\big(1-p_3(t-T_j+T_{j3})\big),
\end{aligned}
\end{equation}
where $m\in\{1,2\}$ such that $j \in S_m$.
We linearize the fluid equations about the equilibrium point given by $i_j^*=q_j^*(i_j^*+d_j^*)$. For each source, $j \in S$, let $u_j(t)$ be an arbitrary perturbation from the equilibrium value of the sending rate such that $u_j(t)=x_j(t)-x_j^*$.
The linearized equations may be written as
\begin{equation}\label{t3_linearized}
\begin{aligned}
T_j\frac{du_j(t)}{dt}=&-a_ju_j(t)-\sum\limits_{k\in S_m}^{}b_{jk}u_k(t-T_j-(T_{km}-T_{jm}))-\sum\limits_{k\in S}^{}c_{jk}u_k(t-T_j-(T_{k3}-T_{j3})).
\end{aligned}
\end{equation}
\subsection{Intermediate drop-tail buffers}
In the case of an intermediate drop-tail buffer, the packet drop probability is given by \eqref{p_d_i}, and the coefficients in \eqref{t3_linearized} are obtained as follows:
\begin{equation}
\begin{aligned}
    a_j=&-x_j^*i_j'+x_j^*q_j^*(i_j'+d_j') \hspace{0.5 cm}\textnormal{for } j \in S,\\b_{jk}=&\frac{x_j^*i_j^*(1-q_j^*)}{q_j^*y_m^*}\hspace{2 cm}\textnormal{for } j,k \in S_m,\\ \textnormal{and} \hspace{0.2 cm}c_{jk}=&\frac{x_j^*i_j^*(1-q_j^*)}{q_j^*y_3^*}\hspace{2 cm}\textnormal{for } j\in S_m, k \in S.
\end{aligned}
\end{equation}
The loop transfer function (return ratio) for the topology is given by 
\begin{equation}\label{L_s_i_2}
\begin{aligned}
    L(s)=&\diag\Biggl(\frac{e^{-sT_j}}{a_j+sT_j}\Biggr) \diag\biggl(\frac{(1-q_j^*)i_{j}^*}{q_j^*}\biggr)\diag(x_j^*)\hspace{0.1cm} R^T(-s) \hspace{0.1cm} \diag \Biggl(\frac{1}{y^*}\Biggr) R(s).
    \end{aligned}
\end{equation}
Proceeding with the control-theoretic analysis, we get sufficient conditions for stability as 
\begin{equation}\label{sc_2_i}
\begin{aligned}
    \frac{i_j(w_j(t))(1-q_j^*)}{q_j^*}&<\frac{\pi}{4} \hskip 2em  \textnormal{for } j \in S.
\end{aligned} 
\end{equation}
These conditions are similar to those of a single bottleneck topology, except the stability margin is reduced by 50\%. This reduction stems from the fact that every source in the topology encounters two bottleneck links. For small values of $q_j^*$, it is too difficult to satisfy these conditions.
\subsection{Small drop-tail buffers}
Since the packet drop probability is given by \eqref{p_d}, the coefficients for small drop-tail buffers are obtained as follows:
\begin{equation}\label{taylor_coeff}
\begin{aligned}
    a_j=&-x_j^*i_j'+x_j^*q_j^*(i_j'+d_j') \hspace{0.3 cm}\textnormal{for } j \in S,\\
    b_{jk}=& \frac{x_j^*i_j^*}{q_j^*}\Biggl(\frac{B_mp_m^*}{y_m^*}(1-p_3^*)\Biggr)\hspace{0.2cm}\textnormal{for } j,k \in S_m,\\
 \textnormal{and}\hspace{.1cm} c_{jk}=& \frac{x_j^*i_j^*}{q_j^*}\Biggl(\frac{B_3p_3^*}{y_3^*}(1-p_m^*)\Biggr)\hspace{0.3cm}\textnormal{for } j \in S_m, k \in S.
    \end{aligned}
\end{equation}
Let $B_1=B_2=B_3=B$. The  return ratio for the topology is given by
\begin{equation}
    \begin{aligned}
        L(s)=&\diag\Biggl(\frac{e^{-sT_j}}{a_j+sT_j}\Biggr)\hspace{0.1cm}\diag(Bi_j^*)\hspace{0.1cm}\diag(x_j)\hspace{0.1cm}Z,
    \end{aligned}
\end{equation}
where the matrix $Z$ is defined as 
\begin{equation}
    \begin{aligned}
        Z&=QR^T(-s)P\hspace{0.1cm}Y^{-1}R(s).
    \end{aligned}
\end{equation}
The matrices $Q$, $P$ and $Y^{-1}$ are defined as follows:
\begin{equation}
    \begin{aligned}
        Q=&\diag\Biggl(\frac{1-q_j^*}{q_j^*}\Biggr),
    \end{aligned}
\end{equation}
\begin{equation}
    \begin{aligned}
        P=&\diag\Biggl(\frac{p_l^*}{1-p_l^*}\Biggr),
    \end{aligned}
\end{equation}
and
\begin{equation}
    \begin{aligned}
        Y^{-1}=&\diag\Biggl(\frac{1}{y_l^*}\Biggr).
    \end{aligned}
\end{equation}
We proceed to find an upper bound for the eigenvalues of $\diag(x_j^*)\hspace{0.1cm}Z$. The spectral radius is given by
\begin{equation}
    \begin{aligned}
\rho(\diag(x_j^*)Z) &=\rho(QR^T(-s)P\hspace{0.1cm}Y^{-1}R(s)\hspace{0.1cm}\diag(x_j^*))\\
    &\leq  \vert | QR^T(-s)P\vert| _\infty \hspace{0.1cm}\vert |Y^{-1}R(s)\hspace{0.1cm}\diag(x_j)\vert |_\infty\\
    &\leq 1.
    \end{aligned}\end{equation}
By similar arguments presented in the single bottleneck scenario, we see that if each source $j$ maintains $i_j(x_j(t))<\frac{\pi}{2B}$, then the system will be stable.
\section{Topology 4: Network with an arbitrary number of bottleneck links}
In general, packets may encounter an arbitrary number of bottleneck links along their route. In such cases, the feedback received by each user from the network will be a measure of aggregate loss probability \eqref{eqn2}. Let $L$ be the set of all links and $S$ be the set of all users in the network. Let $R_j$ be the set of bottleneck links along the path of user $j$, $R_j\subset L$. We assume that $R(s)$ is of full row rank. As mentioned in section \RomanNumeralCaps{4} of this paper, this assumption is reasonable as we consider only bottleneck links in $R(s)$, and in the cases where the same flows traverse through a series of links, only the link with the lowest capacity will act as a bottleneck in equilibrium \cite{Paganini}. The expression for the return ratio and the subsequent analysis follow from the previous section.
\subsection{Intermediate drop-tail buffers}
In the case of intermediate buffers, the sufficient conditions for local stability are given by 
\begin{equation}\label{sc_2_i}
\begin{aligned}
    \frac{i_j(w_j(t))(1-q_j^*)}{q_j*}&<\frac{\pi}{2N_j} \hskip 2em  \textnormal{for } j \in S,
\end{aligned} 
\end{equation}
where $N_j$ is the number of bottleneck links traversed by $j^{th}$ flow. We observe that, as we move from a small buffer regime to an intermediate buffer one, we encounter tighter stability bounds across all network topologies.   
\subsection{Small drop-tail buffers}
Following the same procedure, we see that if $i_j^*<$\scalebox{0.7}{$\dfrac{\pi}{2B}$}  for all $j\in S$, then the entire system will be stable in the case of small drop-tail buffers. Hence, sufficient conditions for the local stability of the system in the presence of an arbitrary number of bottlenecks are 
\begin{equation}\label{eq_51}
    i_j(w_j(t))<\frac{\pi}{2B} \hskip 2em  \textnormal{for } j \in S.
\end{equation}
With intermediate drop-tail buffers, restrictions on window growth functions \big($i_j(x(t))$\big) become more severe as the number of bottleneck links increases, especially with small values of packet-drop probability. Moreover, TCP sources are unaware of the number of bottleneck links traversed by their packets. Hence, the sufficient conditions we obtained are not decentralized. We will need additional signaling to estimate $N_j$. However, with small drop-tail buffers, sufficient conditions for stability are decentralized as they are independent of the number of bottleneck links.
\begin{table}[h]
\small
    \captionsetup{justification=justified} 
     \caption{\textnormal{Scalable sufficient conditions for local stability of networks with $N$ bottleneck links. For small drop-tail buffers, the conditions are decentralized. For intermediate buffers, we need additional signaling for the sources to estimate $N$.}}
    \centering
\begin{tabular}{|p{3cm}|>{\centering\arraybackslash}p{3cm}|>{\centering\arraybackslash}p{4cm}|}
    \hline
 \centering \vspace{0.3cm}TCP Variant & \multicolumn{2}{c|}{\vspace{-0.6cm}}\\ 
     & \multicolumn{2}{c}{Sufficient Conditions \vspace{0.1cm}}\\
    \cline{2-3}
   & \centering{\vspace{0.05cm}Small drop-tail buffers\vspace{0.2cm}} &\vspace{0.05cm}Intermediate drop-tail buffers \vspace{0.2cm}\\ 
   \hline
   \vspace{0.1cm} Generalized TCP & \vspace{0.1cm} $i_j(w_j(t))<\dfrac{\pi}{2B}$ & \vspace{0.1cm}$i_j(w_j(t))<\dfrac{\pi}{2N_j}\dfrac{q_j^*}{(1-q_j^*)}$\\
   \vspace{0.1cm} TCP Reno &  \vspace{0.1cm} $\dfrac{1}{w_j}<\dfrac{\pi}{2B}$&\vspace{0.1cm}$\dfrac{1}{w_j}<\dfrac{\pi}{2N_j} \dfrac{q_j^*}{(1-q_j^*)}$\\ 
     \vspace{0.1cm}Compound TCP & \vspace{0.1cm}$\alpha w_j^{k-1}<\dfrac{\pi}{2B}$ &\vspace{0.1cm}$\alpha w_j^{k-1}<\dfrac{\pi}{2N_j}\dfrac{q_j^*}{(1-q_j^*)}$\\
    \vspace{0.1cm}Scalable TCP &  \vspace{0.1cm}$a<\dfrac{\pi}{2B}$ &\vspace{0.1cm} $a<\dfrac{\pi}{2N_j}\dfrac{q_j^*}{(1-q_j^*)}$\\ \vspace{0.1cm} &\vspace{0.1cm}& \vspace{0.1cm}\\
    \hline
\end{tabular}
\label{conditions_topology_6}
\end{table}
\section{Impact of multiple TCP variants on stability}
Throughout this paper, we have assumed the presence of different variants of TCP, each characterized by the functions $i_j(w(t))$ and $d_j(w(t))$. TCP Reno, Compound TCP, and Scalable TCP are among the TCP algorithms that can be represented using the generalized TCP model. Sufficient conditions for the local stability of TCP Reno, Compound TCP, and scalable TCP are given in Table \ref{conditions_topology_6}. Scalable TCP was designed to outperform traditional TCP variants in high-speed long-distance networks. A sufficient condition for Scalable TCP in a small buffer regime is given by  $a>\frac{\pi}{2B}$, where $a$ is the increment parameter with a recommended value of 0.01. The window increment functions \big($i_j(w_j(t))$\big), and the conditions for local stability are independent of the current window size.\par
We see that sufficient conditions for local stability demand each source to limit its window growth function. This may be interpreted as stability being the shared responsibility of all the competing flows. A set of highly aggressive flows can push the entire system towards instability. 
\section{Impact of delays on stability}
The sufficient conditions we derived do not explicitly contain any delay terms. However, we realize that delays do have an impact on the stability of the system. In the case of TCP, the rate at which each source increases its window size is inversely proportional to the round-trip delay. This is due to the self-clocking mechanism inherent in TCP, where acknowledgments from the receiver trigger the transmission of new packets. This feature is reflected in the exponential term in the return ratio. Consider the case of TCP Reno, where the sufficient condition for stability reduces to $w^*>\dfrac{2B}{\pi}$. Flows with smaller RTT can increase the window size faster. However, for flows with larger RTT, the window growth is slower. If these flows sacrifice too much bandwidth upon a loss (larger $d(w(t))$), then it may lead to instability. Similar arguments can be found in \cite{part_2} as well.
\section{Design guidelines from the necessary condition $a_j>0$}
The necessary condition that must be satisfied by each source to ensure network stability is $a_j>0$ \eqref{taylor_coeff}. Using the expression for $a_j$,
 \begin{eqnarray}
 \begin{aligned}
        -x_j^*i_j'+x_j^*q_j^*(i_j'+d_j')&>0.\\
   \end{aligned}
\end{eqnarray}
If we can choose $i(w(t))$ such that $i_j'<0$, then
\begin{equation}
    \begin{aligned}
        x_j^*|i_j'|+x_j^*q_j^*( d_j'-|i_j'| )&>0\\
       |i_j'| + q_j^*( d_j'-|i_j'| )&>0\\
       d_j' &>|i_j'|-\frac{|i_j'|}{q_j^*}.\\
       \end{aligned}
\end{equation}
Since $q_j^*<1$, it is sufficient to satisfy 
\begin{equation}
    d_j' \geq 0.
\end{equation}
Hence we may choose 
\begin{equation}
   i(w(t))= \alpha(w(t))^{-m} \hspace{2cm} \alpha, m>0 
\end{equation}
and 
\begin{equation}
 d(w(t)) = \beta(w(t))^n \hspace{2.2cm}\beta,n>0.   
\end{equation}
While choosing $\alpha$ and $m$ we should ensure that the sufficient conditions for local stability are satisfied. We can choose $\beta$ and $n$ such that TCP will not lose too much bandwidth while competing with more aggressive or non-responsive flows.
\section{Contributions}
We presented design considerations based on stability for a class of TCP algorithms. We derived sufficient conditions for the local stability of networks operating a generalized TCP algorithm in the presence of heterogeneous round-trip delays. Our model accommodates the simultaneous presence of different TCP variants. We only assume that each flow uses one of the many variants of a generalized TCP model. We explicitly specified sufficient conditions for three TCP variants that come under generalized TCP: TCP Reno, Compound TCP, and Scalable TCP. The conditions we derived are scalable across network topologies with one, two, and many bottleneck links. We considered networks with intermediate and small drop-tail buffers as they offer smaller queueing delays.  The small buffer regime is more attractive as the conditions are decentralized. As we accommodate heterogeneous round-trip delays, multiple bottlenecks, and multiple variants of TCP in our framework, the insights we gather are relevant to the design of new TCP algorithms. TCP algorithms that follow our design considerations can offer stable operation on any network topology, irrespective of the number of bottleneck links or round-trip delays in the network.
\subsection{Avenues for further research}
We intend to carry out extensive packet-level simulations to verify our analytical insights. Also, the dynamics of the networks when multiple TCP variants are present need to be explored. One can also attempt to optimize TCP window functions to realize better specific performance metrics such as throughput, RTT fairness, etc.
\bibliographystyle{IEEEtran}
\bibliography{refs}
\appendix
Proof: TCP Reno, Compound TCP, and Scalable TCP satisfy $a_j\ge 0$.\\
We have 
\begin{equation}
\begin{aligned}
    a_j&=-x^*i'+x^*q^*(i'+d')\\
    &=x^*q^*d'+x^*(1-q^*)(-i')
\end{aligned}
\end{equation}
Hence, to prove $a_j \ge 0$, it is sufficient to prove that $d' \ge 0$ and $i' \le 0$ (since $x^*, q^*, (1-q^*)$ are all positive).\\
\begin{enumerate}[(a)]
    \item TCP Reno:\\
    For TCP Reno, the window update functions are given by
    \begin{equation}
      i(w)=\dfrac{1}{w} \textnormal{ and } d(w)=\beta w  
    \end{equation} where $\beta=\dfrac{1}{2}$. These functions may be expressed in terms of average sending rate $x$ as    
\begin{equation}
  i(x)=\dfrac{1}{xT} \textnormal{  and  } d(x)=\beta xT.  
\end{equation} We get $i'=\dfrac{-1}{T(x^*)^2}$ and $d'=\beta T=\dfrac{T}{2}$. Clearly, $i'<0$ and $d'>0$.\\   
    \item Compound TCP:\\
    For Compound TCP, $i(w)=\alpha w^{k-1}$ and $d(w)=\beta w,$ where the default value of $\alpha, \beta$ and $k$ are given by $\dfrac{1}{8}$, 0.5 and 0.75 respectively. We have    $i(x)=\dfrac{1}{8}(xT)^{1/4}$ and $i'(x)=\dfrac{-1}{32T^{1/4}{x^{*5/4}}}$.  Clearly $i'<0$. We have $d(x)=\beta Tx$ and $d'(x)=\beta T$, which shows that $d'>0$.\\
    
    \item Scalable TCP:\\
    For Scalable TCP, $i(x)=a$, where $a$ is the protocol parameter with a default value of 0.01. Hence, $i'=0$. Also, $d(x)=\beta xT$, which gives us $d'=\beta T$, and hence, $d'>0$.\\
\end{enumerate}
    Thus all three TCP variants satisfy the condition $a_j\geq 0$.
\end{document}